\documentclass[11pt,twoside]{article}
\usepackage{asp2004}
\usepackage{epsf}
\usepackage{psfig}
\usepackage{lscape}

\begin{document}

\title{A Magnetic Rotator Wind-Disk Model for Be Stars}
\author{ M. Maheswaran}
\affil{Department of Mathematics, University of Wisconsin-Marathon
County\\518 S. 7th Avenue, Wausau, WI 54401}

\begin{abstract}
We consider a Magnetic Rotator Wind-Disk(MRWD) model for the
formation of Keplerian disks around Be stars. Material from low
latitudes of the stellar surface flows along magnetic flux tubes
and passes through a shock surface near the equatorial plane to
form a \textit{pre-Keplerian} disk region. Initially, the density
in this region is small and the magnetic field helps to maintain
super-Keplerian rotation speeds. After a fill-up time, the density
of the disk is significantly larger and the magnetic force in this
region becomes negligible compared with the centrifugal force. The
material then moves outwards to form a quasi-steady Keplerian
disk. During the fill-up stage, the meridional component
$B_{m,\star}$ of the magnetic field at the stellar surface must be
larger than a minimum value $B_{m,\star,min}$. The radial extent
of the quasi-steady Keplerian region will be larger when
$B_{m,\star}$ is larger or when viscosity plays a role. In B-type
stars, the values of $B_{m,\star,min}$ are of order 1 G to 10 G.
We find that a condition for the formation of shock-compressed
disk regions is that in faster rotating stars, the wind speed must
be correspondingly larger.
\end{abstract}
\thispagestyle{plain}

\section{Disk Formation Processes}\label{sec-processes}
The MRWD (Magnetic Rotator Wind-Disk) model for the formation of
Keplerian disks involves two processes. The first is a fill-up
process in which material from a star flows along magnetic field
lines and passes through a shock surface to form a pre-Keplerian
disk region. The magnetic field supplies angular momentum to the
wind and assists the flow of material towards the disk region.
Initially, the density in the pre-Keplerian region is small and it
increases significantly during this process.

The second process occurs after the magnetic force in the disk
region becomes small compared to the centrifugal force. Then,
super-Keplerian material in the pre-Keplerian region expands to
form a quasi-steady Keplerian disk. If there is no viscosity or
inflow into the disk region, angular momentum will be conserved.
The time at which the second process starts depends on the
rotational speed, magnetic field strength and rate of fill-up of
the disk region. In some stars, sub-Keplerian wind material may
continue to flow into the super-Keplerian disk region. Also, as
the angular velocity of the expanding disk approaches a Keplerian
distribution, magnetorotational instability (MRI) and viscosity
can become important. When there is inflow or viscosity, the
radial extent of the quasi-steady Keplerian disk region will be
larger. In situations where the magnetic force in the disk is
stronger than the centrifugal force at the end of the first
process, the disk will continue to maintain its pre-Keplerian
structure.

\begin{figure}[!ht]
\begin{center}
\plotfiddle{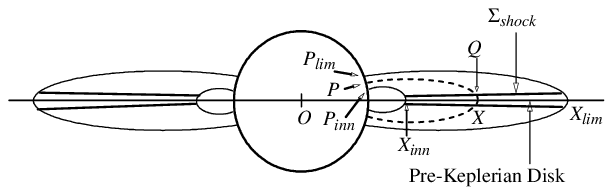}{1.2in}{}{200}{200}{-177pt}{115pt}
\caption{Schematic meridional cross-section of a pre-Keplerian
disk during the fill-up stage. \label{fig1}}
\end{center}
\end{figure}

Figure \ref{fig1} gives a schematic meridional cross-section of a
pre-Keplerian disk during the fill-up stage. It shows the magnetic
field lines from the stellar surface to the inner and outer end
points $X_{inn}$ and $X_{lim}$, respectively, of the pre-Keplerian
region. In a meridional plane, streamlines of flow from the star
to the disk coincide with magnetic field lines. The
\emph{limiting} streamline from $P_{lim}$ to $X_{lim}$ separates
streamlines that flow into the disk region from the open
streamlines that flow away from the star and the disk.
Let $X$ be a point on the equatorial plane in the pre-Keplerian
region. Let $P$ be the point on the stellar surface that is
connected to $X$ by a magnetic field line that loops back to a
point $P\,'$ on the star. If  $\omega$ is the angular velocity at
$P$ and $P\,'$, the angular velocity at $X$ during the
pre-Keplerian stage will also be equal to $\omega$ \citep{mah03}.
The ratio of the angular velocity at $P$ to the critical angular
velocity at the same point is $\alpha = \omega R^{3/2}/
\left[GM(1-\Gamma)\right]^{1/2}$, where $M$ and $R$ are the
stellar mass and radius, respectively. $\Gamma$ is the ratio of
the continuum radiation force to gravity. For points on the
equatorial plane we put $x = r/R$ where $(r, \theta, \phi)$ are
spherical polar coordinates. Let $X_{kep}$ and $X_{esc}$,
respectively, be the points in the pre-Keplerian region at which
the rotational velocities are equal to the Keplerian and escape
velocities. Then, $x_{kep} = \alpha^{-2/3}$ and $x_{esc} = 2^{1/3}
\alpha^{-2/3}$.

\begin{figure}[!ht]
\plotfiddle{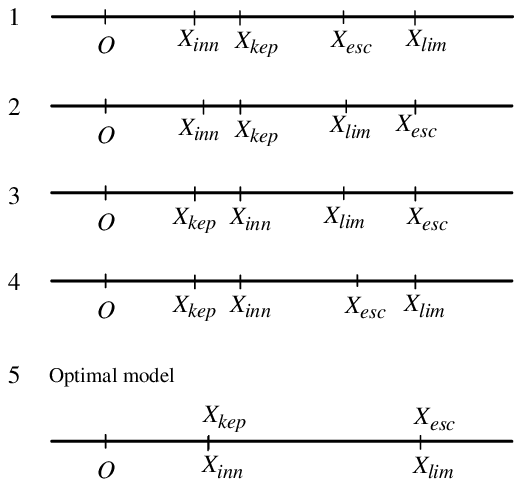}{2.8in}{}{150}{150}{-120pt}{215pt} \caption{ A
schematic picture of the different possible locations of the end
points $X_{inn}$, $X_{lim}$ of pre-Keplerian disks in relation to
the critical points $X_{kep}$, $X_{esc}$.} \label{fig2}
\end{figure}

When the meridional component $B_{m, \star}$ of the magnetic field
at the point $P_{lim}$ on the stellar surface is increased, the
value of $x_{lim}$ also increases. The MRWD model requires that a
super-Keplerian region be present in the pre-Keplerian disk.
Hence, $X_{lim}$ must be further away from the star than $X_{kep}$
so that $x_{lim} > x_{kep}$. Thus, $B_{m, \star}$ must be larger
than a minimum value $B_{m, \star, min}$, which is the meridional
field strength when $X_{lim}$ coincides with $X_{kep}$. Figure
\ref{fig2} schematically shows the different possible locations of
$X_{inn}$ and $X_{lim}$ in relation to $X_{kep}$ and $X_{esc}$.
During the second process, material between $X_{kep}$ and
$X_{esc}$ will move into Keplerian orbits. When there is no
viscosity or inflow of wind material, the material beyond
$X_{esc}$ will flow away. If viscosity or inflow becomes
important, material in the region between $X_{esc}$ and $X_{lim}$
can become part of a quasi-steady Keplerian disk. A pre-Keplerian
region having configuration 1 in Figure \ref{fig2} will yield a
quasi-steady Keplerian disk region with maximal radial extent. The
\emph{optimal model} corresponds to configuration 5, where
$X_{lim}$ coincides with $X_{esc}$ and $X_{inn}$ coincides with
$X_{kep}$ or is closer to the stellar surface than $X_{kep}$. This
model requires the least magnetic field strength at the stellar
surface to form a Keplerian disk with the largest radial extent in
the absence of inflow or viscosity.

The Keplerian region formed when $B_{m, \star}$ is only slightly
larger than $B_{m, \star, min}$ will be in the shape of a ring.
For larger values of $B_{m, \star}$ the radial extent of the
Keplerian region will be larger. In an inviscid model without
inflow, the Keplerian region has its largest radial extent when
$B_{m, \star}$ equals $B_{m, \star, opt}$, which is the field
strength required for the optimal model. For values of $B_{m,
\star}$ larger than $B_{m, \star, opt}$ there is no increase in
the radial extent of the quasi-steady Keplerian region as long as
the material beyond $X_{esc}$ moves away from the disk region.

When the rotation rate $\alpha$ of the sector $P_{lim}P_{inn}$ is
approximately constant, the inner radius of a quasi-steady
Keplerian disk is $\alpha^{-2/3}R$. Let $X_{end}$ be a point on
the outer boundary of the quasi-steady Keplerian disk. When there
is no viscosity or inflow, we have $x_{end} = x_{lim}^4 \;
\alpha^2$ when $B_{m, \star, min} < B_{m, \star} < B_{m, \star,
opt}$ and $x_{end} = 2^{4/3} \; \alpha^{-2/3}$ when $B_{m, \star}
\geq B_{m, \star, opt}$. However, if there is flow of wind
material into the disk region or if there is an onset of MRI
during the second process and viscosity becomes significant, the
value of $x_{end}$ will be much larger because of a contribution
from material between $X_{esc}$ and $X_{lim}$.

\section{Magnetic Field Strengths and Disk Models}\label{sec-crit-mag-field}

Let $B_{m, \star, circ}$ be the lower bound derived by
\cite{mah88, mah92} for the meridional field to withstand the
effects of meridional circulation near the stellar photosphere and
let $t_{circ}$ be the turnover time of this circulation. Let
$B_{m, \star, mtd}$ be the minimum surface magnetic field required
in the MTD model of \citet{cas02}. For Be stars with realistic
rotation rates we have $B_{m, \star, min} < B_{m, \star, circ} <
B_{m, \star, mtd}$. Also, $t_{fill} \ll t_{circ}$, where
$t_{fill}$ is the fill-up time of the pre-Keplerian disk. Table
\ref{table1} gives the critical values of magnetic fields for the
formation of disks in different stellar models.

\begin{table}[!ht]
\caption{Critical values of $B_{m,\star}$ for MRWD and MTD type
disks in different stellar models with specified rotation rates
\citep{cas02,mah03}} \label{table1}
\begin{center}
\small{
\begin{tabular}{llccccr}
\tableline\\
 & & \multicolumn{2}{c}{MRWD Model} & & & MTD Model\\
 \cline{3-4}  \cline{7-7} \\
  Spectral & $\alpha$ & $B_{m,\star,min}$ & $B_{m,\star,opt}$ & $\quad B_{m,\star,circ}$
  & $t_{circ}$  & $B_{m,\star,mtd}$  \\
   Type & & (G) & (G) & \quad (G) & (yr) & {(G)\quad} \\
  \tableline\\
  O3   & 0.6  & $437\quad$   & $893\quad$  & \qquad$139\quad$ & $0.53$  & 22000  \\
  O6.5 & 0.6  & $101\quad$   & $207\quad$  & \qquad$175\quad$ & $0.57$  & 5000  \\
  B0   & 0.6  & $33 \,\,\,$  & $67\,\,\,$  & \qquad$131\quad$ & $0.31$  & 1600  \\
  B2   & 0.5  & $7\,$        & $13\,\,\,$  & \qquad$25\,\,\,$ & $1.1\,\,$   & 335  \\
  B5   & 0.45 & $\;\,1.4$   & $\;\,2.5$    & \qquad$\;\,3.6$  & $6.7\,\,$   & 64  \\
  B9   & 0.4  & $\;\,0.8$   & $\;\,1.3$    & \qquad$\;\,0.5$  & $30.6\,\,\,$  & 30  \\
  \tableline
\end{tabular}
}
\end{center}
\end{table}

Here, we consider models that are appropriate for stars with
different values of $B_{m,\star}$. Although the azimuthal
component may also play a role, we do not include it in this
discussion.

(a) If $B_{m,\star,min}<B_{m,\star}<B_{m,\star,circ}$,
quasi-steady Keplerian disks satisfying the MRWD model can be
formed. The surface magnetic field will be affected by meridional
circulation near the photosphere. This can lead to changes in the
disk over time scales of order $t_{circ}$.

(b) When $B_{m,\star,circ}<B_{m,\star}<B_{m,\star,mtd}$, the MRWD
model is appropriate. The surface magnetic field will be able to
withstand the effects of circulation and persist over time periods
that are long compared with $t_{circ}$.

(c) If $B_{m,\star} > B_{m,\star,mtd}$, the disks will be of MTD
type. If the magnetic field lines from the star thread the disk
region and loop back to the star, like field lines of a
dipole-type field, the angular velocity distribution in this disk
will be the same as that in the region $P_{inn}P_{lim}$ of the
stellar surface from which the field lines emerge. On the other
hand, if the field lines from the star pass through the disk
region and travel outwards to large distances from the star
without looping back, the rotation law for the disk will be
similar to the empirical formula specified in the MTD model of
\citet{cas02}.

\section{Discussion}\label{sec-discussion}

The requirement that the disk density should be positive gives the
condition $\, v^2_n  > \left({B^2_{disk} -
{B}^2_{wind}}\right)/{8\pi \rho_{wind}} \,$ for disk formation
\citep{mah03}. Here, all quantities are evaluated on the
appropriate sides of a point $Q$ on $\Sigma_{shock}$ and $v_n$ is
the normal component of the wind velocity. Thus, for given values
of the equatorial rotation speed and surface magnetic field
strength, the wind velocity must be larger than a critical value
for a disk to be formed. Since this critical value depends on the
stellar rotation speed, a faster rotation rate does not ensure the
formation of a shock-compressed disk region unless $v_n$ is
correspondingly larger.

The MRWD model does not require that the magnetic field be
uniformly strong across the entire stellar surface. The field must
have the required strength on the sector $P_{inn}P_{lim}$, whose
spread in latitude is only a few degrees. The basic processes in
this model do not depend on whether the field is axially
symmetric. The results obtained for symmetrical systems will
qualitatively apply to stars possessing magnetic fields with flux
loops that thread the disk region.

Observational evidence \citep[e.g.,][]{don01,don02,nei02,nei04}
indicates that several stars of the type we consider for disk
formation are oblique magnetic rotators with dipole fields.
\citet{pre04} find that disk-like structures can form along the
magnetic equatorial planes of models with strong magnetic fields
in which the Alfv\'en speed is faster than the rotation speed of
the magnetosphere. Thus, when the MRWD model is applied to oblique
rotators, a pre-Keplerian region can be formed along a plane or
surface that is determined by gravity, centrifugal force and
magnetic force acting on the disk material. During the second
process, if the magnetic force in the disk becomes small, we
expect that super-Keplerian material will move into Keplerian
orbits in the rotational equatorial plane. In the case of oblique
rotators with $ B_{m,\star} > B_{m,\star,mtd}$, the results of
Preuss et al imply that an MTD disk will be located in the
magnetic equatorial plane rather than the rotational equatorial
plane.

Because the magnetic force in the disk region of the MRWD model
becomes small at the end of the fill-up stage, the one-armed
spiral pattern of the Global Disk Oscillation model
\citep[e.g.,][]{oka97} can be used to explain the V/R variability
in disks of Be stars.

\acknowledgements I wish to thank the University of Wisconsin
Marathon County Foundation for a 2003 Summer Grant.

{}

\end{document}